\documentclass[%
 reprint,
 amsmath,amssymb,
 aps,
pre,
floatfix,
]{revtex4-1}

\usepackage{graphicx}% Include figure files
\usepackage{dcolumn}% Align table columns on decimal point
\usepackage{bm}% bold maths
\usepackage{epstopdf}
\usepackage{siunitx}
\usepackage{gensymb}
\usepackage{float} % Enable placing pictures elsewhere than top page
\begin{document}

%\preprint{APS/123-QED}

\title{Hippopede curves for modelling radial spin waves in an azimuthal graded magnonic landscape }% Force line breaks with \\

\author{D. Osuna Ruiz *\textsuperscript{1}} 
\email{do278@exeter.ac.uk}
% \altaffiliation[Also at ]{Department of Physics and Astronomy, University of Exeter.}%Lines break automatically or can be forced with \\
\author{A. P. Hibbins\textsuperscript{1}}
\author{F. Y. Ogrin\textsuperscript{1}}%
 
\affiliation{%
 \textsuperscript{1}Department of Physics and Astronomy, University of Exeter, Exeter EX4 4QL, United Kingdom.
 }%

\date{\today}% It is always \today, today,

\begin{abstract}
We propose a mathematical model for describing radially propagating spin waves emitted from the core region in a magnetic patch with n vertices in a magnetic vortex state. The azimuthal anisotropic propagation of surface spin waves (SSW) into the domain, and confined spin waves (or Winter's Magnons, WM) in domain walls increases the complexity of the magnonic landscape. In order to understand the spin wave propagation in these systems, we first use an approach based on geometrical curves called \lq hippopedes', however it provides no insight into the underlying physics. Analytical models rely on generalized expressions from the dispersion relation of SSW with an arbitrary angle between magnetization \textbf{M} and  wavenumber \textbf{k}. The derived algebraic expression for the azimuthal dispersion is found to be equivalent to that of the \lq hippopede' curves. The fitting curves from the model yield a spin wave wavelength for any given azimuthal direction, number of patch vertices and excitation frequency, showing a connection with fundamental physics of exchange dominated surface spin waves. Analytical results show good agreement with micromagnetic simulations and can be easily extrapolated to any n-corner patch geometry. 

\end{abstract}

\pacs{Valid PACS appear here}% PACS, the Physics and Astronomy
\maketitle

%\tableofcontents

\section{\label{sec:level1}introduction}

Due to their low loss and shorter wavelength compared to electromagnetic waves in free space, spin waves are a promising candidate for comunicating information in micron and sub-micron scale magnonic circuits \cite{0022-3727-43-26-264001,Karenowska2016,PhysRevApplied.4.047001}. Spin wave spectra of magnetic circular nanodots have been studied intensively \cite{PhysRevB.93.184427, doi:10.1063/1.1772868, PhysRevB.76.224401, PhysRevB.73.104424}. When an in-plane magnetic field excitation is applied to a vortex spin configuration, the lowest energy mode that can be excited is the gyration of the vortex core, which depends upon the aspect ratio of the disc \cite{YuCore}. At higher frequencies, higher order gyrotropic modes \cite{higher, Guslienko2015GiantMV} and a complete set of modes related to azimuthal and radial spin waves appear \cite{PhysRevB.93.184427}. The latter type of spin waves are related to Damon-Eshbach modes where $\mathbf{k}$ is perpendicular to $\mathbf{M}$ in a vortex core configuration \cite{PhysRev.118.1208}, and their spectra is strongly dependent on thickness and more generally, on the physical geometry of the patch. Moreover, magnetization inhomogeneities such as vortex cores have attracted attention as spin waves emitters \cite{PhysRevLett.122.117202,article}. It is known that, due to the confinement or the natural magnetic state of the sample, inhomogeneities of the internal magnetic field can be sources of spin waves due to a graded index in the magnonic landscape \cite{PhysRevB.96.064415,doi:10.1063/1.4995991}.  Spiralling spin waves found in vortex configurations have been explained as hybridization of stationary azimuthal spin waves and higher order gyrotropic modes, therefore showing no radial propagation \cite{PhysRevLett.122.117202,10.3389/fphy.2015.00026,Kammerer2011MagneticVC}. 

\par Spin waves with spiral or circular wavefronts have been reported through micromagnetic simulations and experiments in simple elements such as circular discs or square patches \cite{article,PhysRevLett.122.117202,PhysRevB.100.214437}. In Ref. \cite{PhysRevLett.122.117202}, the authors propose an analytical expression for the dispersion relation of radially propagating exchange-dominated spin waves from the core region, which are explained as laterally emitted spin waves from a first order gyrotropic mode of the vortex core. Therefore, they manifest a 'Surface Spin Wave -like' ('SSW') propagating behavior (since \textbf{M} is perpendicular to \textbf{k}). 

\par In the past, a vast work on analytical modelling has dealt with magnetically non-saturated structures presenting domain walls. Some examples are spin wave emission from Bloch domain walls \cite{PhysRevB.96.064415}, reflection and transmission across domain walls \cite{article_analyticalblochspinwave,doi:10.1063/1.4799285} or magnetic configuration in a transition between domain walls types \cite{PhysRevB.92.214420}. These models help to provide insight into the dynamics and a tool for modelling spin wave phenomena in confined structures that show a more complex magnonic landscape than saturated dots. To the best of our knowledge, we have not found in the literature a generalized mathematical model dedicated to the particular physics of the spiral or circular spin waves emitted from a point source in confined structures of more complex geometries than a circular disc, which are expected to considerably reshape the radial wavefront \cite{PhysRevB.100.214437}.

\par Following on these studies, in this work we report on a model for the observed wavefront of spin waves emitted from an almost point source (e.g., a vortex core) in any n-vertex patch, which implies the existence of domain walls, azimuthally distributed across the geometry. The final expression is derived from a generalisation of the dispersion relation of surface waves for an arbitrary angle between magnetisation \textbf{M} and wavenumber \textbf{k}. For n $>$ 2, the patch adopts the form of a regular polygon, for n = 2 and n = 1, the model considers two or one single vertices. Finally, n = 0 implies a circular disc. The obtained curves from the model agree well with numerical results. In section II we describe our models and the numerical method used for their validation. In section III, we provide with a comparative of both methods and their validation through numerical simulations as well as discussion of results. This model can help on the description of complex spin wave wavefronts in non-saturated elements, which avoids running numerical simulations for particular shapes. Due to the attention that the experimentally observed short-wavelength radial spin waves have recently drawn, we believe this model is of interest to researchers, experimentalists mainly, working in the field of spin waves emitted from a point source.

\section{\label{sec:level2}Numerical methods and calculations}

In order to obtain an analytical solution for the spin wave wavefront in a non-saturated patch, we must make some initial approximations. This is due to the remarkable complexity of the magnetic configuration within the patch along the azimuthal direction, specially near magnetically inhomogeneous regions, which are determined by shape anisotropy and dipolar and exchange interactions solely. Thus, in our first approach we mathematically infer a general equation from numerical results and in our second approach, we extrapolate analytical results from a simpler case scenario to ours. Despite of the apparent crudity of these extrapolations, both models show good agreement and therefore considered reliable for, at least, descriptive purposes. However, our second approach is more fundamental and still, even after a crude generalization, it also shows very good agreement with micromagnetic results. 

To obtain more insight into the dynamics and confirm the performance of our model, we performed a set of micromagnetic simulations using Mumax3 \cite{doi:10.1063/1.4899186}. We simulated a circular microdisc with the typical material parameters of permalloy at room temperature with saturation magnetization $M_\text{s}$ $= 8$ $\times$ $10^5$ Am$^{-1}$, exchange constant $A_{\text{ex}} = $ $1.3$ $\times$  $10^{-11}$ Jm$^{-1}$, Curie temperature from a weighted average of iron and nickel $T_{\text{C}}$ $=$ $270$ K and Gilbert damping constant $\alpha$ $=$ $0.008$. With these parameters, the single circular disc was simulated in a hexaedral grid. Shapes with diameter ${d}$ of 900 nm and thickness ${t}$ of 80 nm were simulated. The grid was discretized in the ${x, y, z}$ space into 512 $\times$ 512 $\times$ 16 cells. The cell size along ${x}$ and ${y}$ was 3.9 nm, while the cell size along ${z}$ was fixed to 4 nm. The cell size along three dimensions is always kept smaller than the exchange length of permalloy (5.3 nm). The number of cells was chosen to be powers of 2 for sake of computational efficiency. We also set a \lq smooth edges' condition with value 8 \cite{doi:10.1063/1.4899186}.
A key point in micromagnetic simulations is to achieve a stable equilibrium magnetization state. We first set a vortex state with polarity and chirality numbers of  (1, $-$1) and then executed the simulation with a high damping ($\alpha$ $=$ 1) ) to relax the magnetization until the maximum  torque/$\gamma$ (\lq maxtorque'  parameter  in Mumax3) reached {$10^{-7}$} T indicating convergence and the achievement of a magnetization equilibrium state. The typical time to achieve the equilibrium state was 100 ns. Once the ground state was obtained, damping was set back to $\alpha$ $=$ 0.008 and the relaxation process repeated. The microdisc spin configuration was recorded as the ground state of the sample and  then  used  for the simulations with the dynamic activation. 

For analyzing time evolution of the magnetic signal, we apply a continuous wave excitation at the core region with a magnetic field $B_0$ at a specific frequency {$f_{0}$},

\begin{gather}
B_{0}(t) = A_{0}\text{sin}(2\pi f_{0} t)), 
\end{gather}

where {$f_{0}$} is the microwave excitation frequency and pulse amplitude {$A_{0}$ = 0.3 mT}. This is small enough to remain in the linear excitation regime and avoid any changes to the equilibrium state. A sampling period of {$T_{s} = $25 ps} was used, recording up to 200 simulated samples in space and time, only after the steady state is reached.

\par In the next sections we describe the proposed models and their derivations. Finally, validations for each of them through numerical simulations are shown.

\subsection{First approach} 

For this study, we mathematically infer a fitting model from numerical results on the first obtained shapes when n = 0, 1, 2, 3, 4... and so on. We then generalize it to any n-vertex patch. 
For the case of $n = 1$ we take an internal angle assumed to be $\pi/3$ and only one domain wall is present, resembling the patch to a \lq teardrop' shape. For $n = 2$, we also make a similar assumption and the patch resembles to a \lq double teardrop' shape. For larger values of $n$, the internal angles of the vertices are the internal angles of the regular polygons, defined as,  $2\pi(n-2)/n$. Of course, for  $n = 0$ we have a circle. The vertices are distributed around the shape, separated by $2\pi/n$ radians and the resulting domain walls spaced by $\pi/n$.
\par Regarding magnetic configuration in equilibrium after a relaxation process, and assuming a centred vortex core, $(n-1)$ triangular domains and $n$ domain walls will form in the patch. In contrast to the circular dot, the azimuthal distributed domain walls will distort the wavefront of the propagating spin wave from the core region, introducing an azimuthal dependence (or equivalently, $n-$ dependence) to the wavelength of the radial wave ($\lambda(\theta)$). Also, two known values for the spin wave wavelength can be analytically deduced for any n-corner patch: the characteristic wavelength of an exchange dominated surface spin wave ($\lambda_{\text{SSW}}$), when \textbf{k} is perpendicular to \textbf{M} (this is, when the spin wave propagates into the domain), and the characteristic wavelength of the confined spin wave along the domain wall ($\lambda_{\text{WM}}$), also known as Winter's magnon \cite{PhysRev.124.452}. Fig. 1 shows the characteristic dispersion relations for the laterally emitted spin wave from the vortex core from Ref. \cite{PhysRevLett.122.117202} (blue curve) and the exchange-dominated Winter's magnon in an ideal 180 degrees Bloch wall \cite{PhysRevLett.114.247206} (orange curve). It is worth noting that, in an n-vertex patch, the formed domain walls will be of the angle of the vertex. For example, in a square (n = 4), this is an angle of 90 degrees (see the top-right inset in Fig. 1). Micromagnetic simulations (not shown here) show that the expected wavenumber is reduced with respect to the 180 degrees Bloch wall, due to the shape anisotropy of the sharp corner. This latter study is not in the scope of this article (although it will be addressed in a future work) and the effects of an intermediate domain wall are not included in our model. For practical purposes, we obtain $k_{\text{WM}}$ from the dispersion relation of an ideal 180 degrees Bloch wall (see orange curve in Fig. 1). For excitation frequencies at which both modes coexist, the wavenumbers (or wavelengths) that fall in the grey area, delimited by $k_{\text{SSW}}$ and $k_{\text{WM}}$, can relate to the azimuthal-dependent wavelength of the radial spin wave in the patch.

\begin{figure}[ht]
\centering 
\includegraphics[trim=0cm 0cm 0cm 0cm, clip=true, width=9cm]{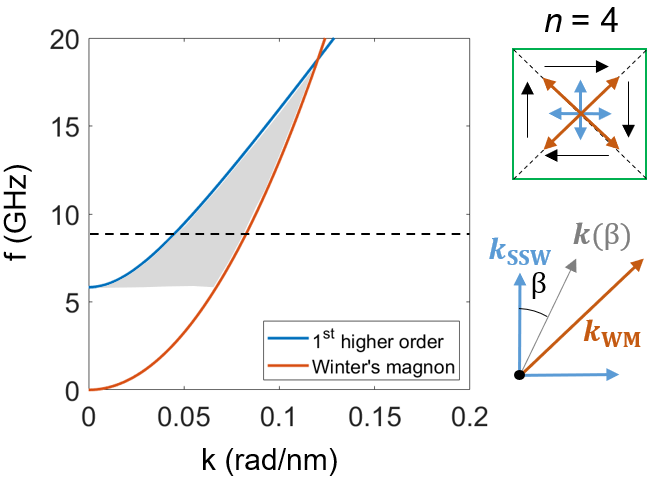}
\caption{Dispersion relations for the laterally emitted spin wave from a first higher order dynamical core, see Ref.\cite{PhysRevLett.122.117202}, and a Winter's magnon (WM), for the material parameters indicated in section II, as limiting cases. Dashed horizontal line indicates an excitation frequency of 8.8 GHz. Grey area highlights the expected magnitudes for intermediate wavevectors (see bottom-right inset) between the two limiting cases, when being simultaneously excited. (Insets) Colour arrows show the wavevectors for the laterally emitted spin wave (in a SSW configuration) and the Winter's magnon, which are assumed $k_{\text{WM}}\approx2k_{\text{SSW}}$ for the excitation frequency of 8.8 GHz (length of the vectors represent their magnitudes). Black arrows show the orientation of magnetisation in the domains. Bottom inset shows a simplified schematic of the angular distribution of wavevector at the top-right corner of a square patch (n = 4). Grey arrow is an intermediate case for the spin wave wavevector between the limiting cases, from what an angular dependence can be inferred.}\label{Fields1}
\end{figure}

\begin{figure}[ht]
\centering 
\includegraphics[trim=0cm 0cm 0cm 0cm, clip=true, width=8cm]{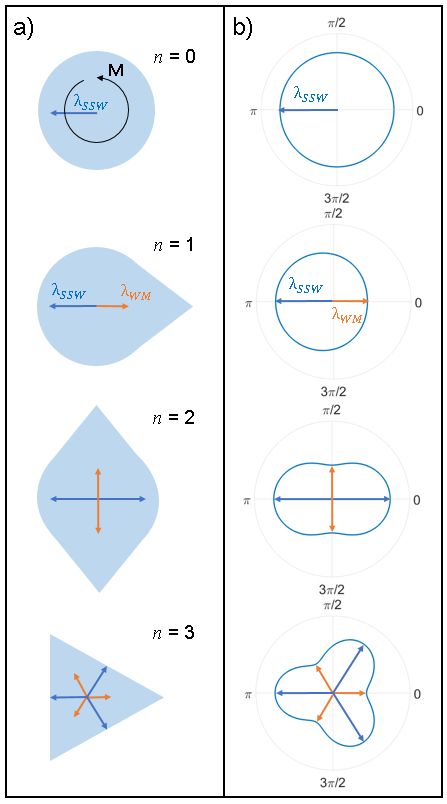}
\caption{(a) Schematic of various magnetic patches depending of the number of vertices $n$ with magnetization around the core (solid black line). Domain walls bisect the sharp corners. (b) Polar representation of Eq. (2)  showing the azimuthal variation of wavelength within the shape, with the appropriate rotation of the patch ($\phi_0$) so the equation describes correctly the shape of the patch. Minimum and maximum radial amplitudes are assumed $\lambda_{\text{SSW}}=2\lambda_{\text{WM}}$ (length of the vectors represent their magnitudes). In terms of the hippopedes parameters $a$ and $b$, this means a ratio of $b/a=0.75$.}  \label{Fields1}
\end{figure}

\par The proposed model is based on the mathematical expressions of a family of curves known as \lq hippopedes'. Since the polar representation of these curves allow a smooth azimuthal transition from a certain wavelength (maximal) to another finite value (minimal), these curves can be used here as a generalization of the problem scenario. By applying these expressions to the particular scenario, i.e. a magnetic patch with regularly distributed vertices and domain walls, a simple expression for the spin wave wavelength can be obtained. From a conic canonical equation with geometrical parameters $a$ and $b$, see Supplemental Material (1) for a more detailed description, then the generic equation of the resulting hippopede in polar coordinates is,
\begin{gather}
f(\theta)=2\sqrt{b}\sqrt{a-b\cdot\text{sin}^2(\theta\frac{n}{2} + \phi_{0})}, 
\end{gather}
where the phase parameter $\phi_{0}$ sets an initial rotation angle for the patch. For $n=2$ and $b=2a$, Eq.(2) leads to a special case known as Bernoulli's Lemniscate. In fact, a whole family of lemniscates can be obtained from the hippopedes if $b>a$. More information on the \lq Hippopede curves' can be found in the Supplemental Material (1). The \lq hippopedes' when $b<a$, known as Booth's ovals, allow a transition from a finite wavelength maximum value ($\lambda_{\text{SSW}}$) to another non-zero minimum value ($\lambda_{\text{WM}}$), see Fig. 2(b). The particular values of the geometrical parameters $a$ and $b$ can be found from the \lq hippopede' general equation particularized to the wavelength limiting conditions of a maximum $\lambda_{\text{SSW}}$ at every $2\pi/n$ angle and a minimum $\lambda_{\text{WM}}$ at every $2\pi/n+\pi/n$ angle. An initial rotation angle of $\phi_{0}=0$ is assumed. The ratio $b/a$ is found to be equal to $1-\lambda_{\text{WM}}^2/\lambda_{\text{SSW}}^2$. In the range of frequencies under study, $\lambda_{\text{SSW}}>\lambda_{\text{WM}}$ is satisfied, so this implies we can model our system with hippopede curves where $b<a$. The initial phase rotation $\phi_{0}$ in Fig. 2(b) (and hereafter) is chosen so it properly coincides with the numerically modelled patch. Therefore, a more complete expression for our model is,

\begin{gather}
\lambda(\theta,n)=\lambda_{\text{SSW}}\sqrt{1-(1-\frac{\lambda_{\text{WM}}^2}{\lambda_{\text{SSW}}^2})\text{sin}^2(\theta\frac{n}{2}+\phi_{0})}. 
\end{gather}
Fig. 2(b) shows a collection of curves from Eq.(3) for different number of vertices $n$. Following from the magnetic configuration of the patch in the vortex state and assuming for example $\lambda_{\text{SSW}}=2\lambda_{\text{WM}}$, which yields a ratio $b/a=0.75$ ($b<a$), results qualitatively show an azimuthal changing wavelength around the centre of the shape according to a hippopede curve. In Section II.B, we derive a more generalized expression that yields a connection with the fundamental physics.

\subsection{Second approach} 

In this case, we start from the analytical expression of the dispersion relation of surface spin waves given an arbitrary angle $\alpha$ between \textbf{M} and \textbf{k}. Damon and Eshbach \cite{PhysRev.118.1208} found a generalised expression for the dispersion relation of surface spin waves in a semi-infinite stripe for arbitrary angles $\alpha$ and $\beta$, being $\alpha$ the angle between the ferromagnetic planar body surface and effective field $H_{i}$, and $\beta$ the angle between the orthogonal direction of that effective field and wavenumber \textbf{k}. Extended to the exchange regime, it can be expressed as,

\begin{gather}
\omega= \frac{\gamma H_{i}}{2\text{cos}\alpha\text{cos}\beta}+ \frac{\gamma B_{i}}{2}\text{cos}\alpha\text{cos}\beta + \omega_{M}\lambda_{\text{ex}}k^2. 
\end{gather}
In Ref. \cite{PhysRev.118.1208}, the original expression is derived for magnetostatic spin waves in a semi-infinite stripe. It should be noted that our problem scenario, although being a finite sample, can still be regarded equivalent due to the short wavelength of the spin wave modes under study \cite{PhysRev.118.1208}. In Ref. \cite{PhysRev.118.1208}, the expression shows a continuous variation of the spin wave wavelength as angle β increases, towards the limiting scenario of a Backward Volume spin wave (BVSW) configuration. It is worth to note that, in our problem scenario, that limiting case would not be the BVSW dispersion relation but the Winter's magnon's (see Fig. 1). Also, in non-saturated samples, Eq. (4) can be reduced by specifying only in-plane magnetization ($\alpha = 0$), assuming the internal field in the magnetic domain is $\textbf{H}_{i} = -\textbf{M}_{S}$ (therefore, $H_{i} =\text{M}_{S}$ and $\textbf{B}_{i} = 0$) in absence of an external biasing field. For $\beta = 0$, \textbf{k} is perpendicular to \textbf{M} (\textbf{k}$\perp$\textbf{M}, and therefore $\lambda=\lambda_{\text{SSW}}$ (see inset in Fig. 1). Hence, for a specific excitation frequency $\omega =\omega_{0}$ we can rewrite Eq. (4) in terms of a variable wavelength ($\lambda=2\pi/k$) in the azimuthal direction $\beta$ (as defined in Fig. 1 and in Ref. \cite{PhysRev.118.1208}) as (a step-by-step derivation is shown in Supplemental Material (2)),
\begin{gather}
\lambda(\beta)=\lambda_{\text{SSW}}\sqrt{\frac{(1-p)\text{cos}\beta}{\text{cos}\beta-p}}, 
\end{gather}
where $\beta=\theta$, being $\theta$ the azimuthal direction as defined in Eq. (2) and Eq. (3) that coincides with the angle $\beta$, setting the reference for an azimuthal dependence of \textbf{k} at $\theta = 0$, and $p=\omega_{M}/2\omega_{0}$, where $\omega_{M}=\gamma M_{s}$. We obtain a classical surface spin wave dispersion behavior from Eq. (5) when $\theta=0$. In Eq. (2), the initial arbitrary rotation phase of the patch can be conveniently chosen as $\phi_{0}=\pi/2$ so a surface spin wave wavelength can be effectively obtained for $\theta=0$, as reference point. This implies we can substitute $\text{sin}(\theta + \phi_{0}) \rightarrow \text{cos}(\theta) $ in Eq. (2), which keeps the reference $\lambda(0)=\lambda_{\text{SSW}}$ consistent with Eq. (5). Eq. (5) implies a decreasing wavelength as the angle $\theta$ (or equivalently, $\beta$) increases. For a flux closure magnetisation in the patch, the reference angle coincides with the direction of propagation into a first magnetic domain. 
\par In the model, parameter $p=\omega_{M}/2\omega_{0}$ yields a connection between the observed radial wavefront in simulations and the magnetic properties of the material but sets an upper frequency bound for the model, which is not physically meaningful. The model from Eq. (5) yields imaginary values for $\text{cos}(\theta) < p\leq 1$ and therefore, it would only be applicable for $\omega_{0} < \omega_{M}/2$ and for certain azimuthal directions. Eq. (5) would still be applicable as a model for the scenario of a magnetic patch for small deviations of $\theta$ from 0, although it does not apply when $\theta\rightarrow \pi/2$, where backward volume spin wave propagates instead of surface spin waves according to Ref.\cite{PhysRev.118.1208}. 
\par Also, this preliminary model assumes in-plane magnetisation for all azimuthal directions, so it still does not take into account effects of magnetic inhomogeneities, i.e., the domain walls. Fig. 3 shows a collection of curves from Eq. (5) illustrating a periodic effect when an n number of corners is included ($\theta \rightarrow n\theta/2 $). In Fig. 3, only the rightmost lobe would be strictly represented by Eq. (5), i.e., for values of $-\pi/n<\theta < \pi/n$ and $n>0$. 

\begin{figure}[ht]
\centering 
\includegraphics[trim=0cm 0cm 0cm 0cm, clip=true, width=8cm]{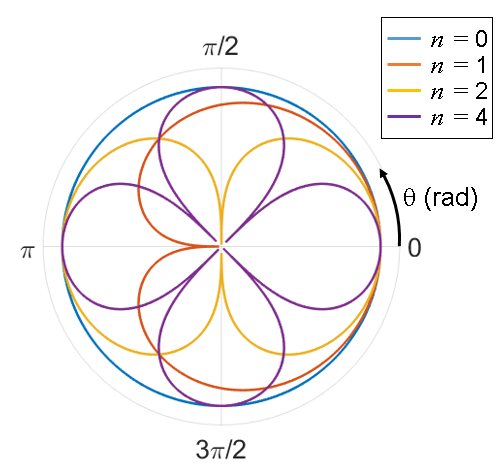}
\caption{Collection of curves from Eq.(5) for $n=0, 1, 2$ and $4$. Curves for $n=3$ and higher orders $(n>4)$ can be easily inferred. Physically non-realizable zeroes are placed at every $\pi/n$ angle.}  \label{Fields1}
\end{figure}

As stated above, imaginary values and zeroes of Eq. (5) due to the frequency dependence and geometry, are not physically meaningful in our scenario. This is due to the imposed lower frequency gap for the FMR in the disc and the presence of domain walls, respectively. To include these phenomena and avoid the zeroes in the model, Eq. (5) has to be generalised with the undefined parameter $\sigma$ and re-scaling factor $\epsilon$ so the expression is extended as,
\begin{gather}
\lambda(\theta,n)=\epsilon\cdot\lambda_{\text{SSW}}\sqrt{\sigma + \frac{(1-p)\text{cos}(n\frac{\theta}{2})}{\text{cos}(n\frac{\theta}{2})-p}}. 
\end{gather}
\par In an n-corner patch, the spin wave shows a wavelength of $\lambda_{\text{SSW}}$ at every $2\pi/n$ angle and of $\lambda_{\text{WM}}$ at every $\pi/n$ angle. These limiting conditions allow to find the values of parameters $\epsilon$ and $\sigma$. The values for $\epsilon$ and $\sigma$ are found to be: $\epsilon =(\sqrt{\lambda_{\text{SSW}}^2-\lambda_{\text{WM}}^2})/\lambda_{\text{SSW}}$ and $\sigma =\lambda_{\text{WM}}^2/(\lambda_{\text{SSW}}^2-\lambda_{\text{WM}}^2)$, which simplified for $\lambda_{\text{WM}}=0$, lead to the equation from Ref. \cite{PhysRev.118.1208}. Due to the azimuthal periodicity every $\pi/n$ (in contrast to the scenario described in Ref. \cite{PhysRev.118.1208}), cosine terms in Eq. (6) must only take positive values. Also, since taking their absolute value would yield a non differentiable function at the angle where the domain wall is encountered, the cosine terms are replaced by their squared values. Assuming $\lambda_{\text{WM}}\neq0$ and after algebraic transformations, the modified equation is,
\begin{gather}
\lambda(\theta,n)=\lambda_{\text{WM}}\sqrt{1-(1-\frac{\lambda_{\text{SSW}}^2}{\lambda_{\text{WM}}^2})\frac{(1-p)}{\text{cos}^2(n\frac{\theta}{2})-p}\text{cos}^2(n\frac{\theta}{2})}, 
\end{gather}
where $p=\omega_{M}/2\omega_{0}$. Our key result is obtained if $p>>1$ and therefore $(1-p)/(\text{cos}^2(n\theta/2)-p)\approx 1$. Then, the resultant equation is indeed the Hippopede curve equation (assumed $\phi_{0}=\pi/2n$, so the directions for $\lambda_{\text{WM}}$ and $\lambda_{\text{SSW}}$ are exchanged) described in Section II.A, which explains the good fitting to these curves at low frequencies ($\omega_{0} << \omega_{M}/2$). Therefore, the azimuthal change of wavelength for a spin wave emitted from the core in an n-corners magnetic patch in the vortex configuration is fully described by Hippopede curves. 

\begin{figure}[ht]
\centering 
\includegraphics[trim=0cm 0cm 0cm 0cm, clip=true, width=9cm]{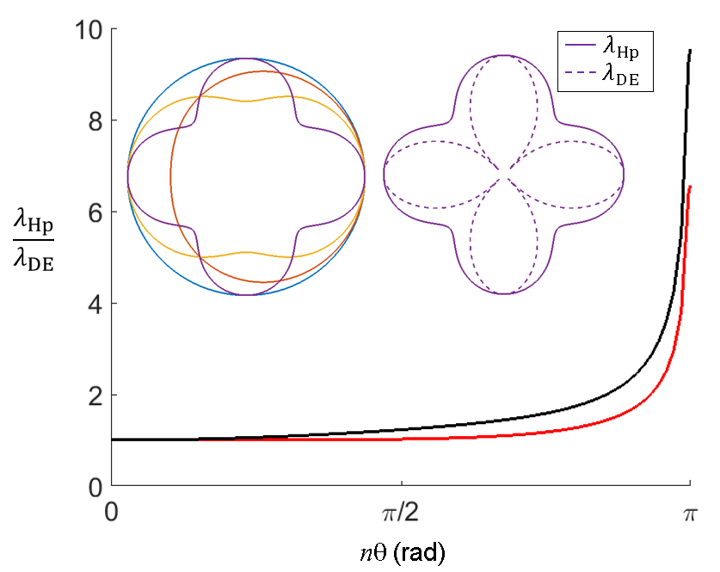}
  \caption{Ratio (black solid curve) between wavelengths from the dispersion relation from \cite{PhysRev.118.1208} ($\lambda_{\text{DE}}$) and the extended model linked to the Hippopedes ($\lambda_{\text{Hp}}$) for $p \approx 1.75$ showing good agreement, and practically no deviation from unity at any angle apart from the domain wall region for $p\approx 10.75>>1$ (red solid curve). Inset (left): Normalised results from Eq. (7) adding azimuthal periodicity for a number of corners $n = 0$ (blue), $n=1$ (red), $n=2$ (yellow) and $n=4$ (purple). Inset (right): Comparison of the normalised results from Eq. (7) (solid purple curve) and Eq. (5) (dashed purple curve) for $n=4$ and $p \approx 1.75$.}  \label{Fields1}
\end{figure}

\par Fig. 4 (left inset) shows a collection of curves obtained from Eq. (7) normalized to $\lambda_{\text{SSW}}$, assuming  $\lambda_{\text{SSW}} = 2 \lambda_{\text{WM}}$ (see Fig. 1). In the right inset, a particular case for $n=4$ is shown in comparison with the respective solution from Eq. (5) and the relative error between both equations for all angles between 0 and $\pi/n$, normalised to the number of vertices $n$ for a value $p\approx 1.75$ (black solid curve), obtained from assuming $\omega_{0}/2\pi = 8.8$ GHz and from Permalloy material properties, $\omega_{M}/2\pi \approx 27$ GHz. For a large enough value of $p$, assumed to be about six times larger ($p\approx 10.75$), and therefore $p>>1$, results show indeed a minimal difference with the hippopede curves (red solid curve), almost negligible away from the domain wall regions. The proposed model, as an extension of Eq. (5), avoids the azimuthal zeroes and shows a minimal difference with the values from the exchange dominated \lq surface spin wave' dispersion. Therefore, Eq. (7) is a suitable model, derived as a generalization from Eq. (5). In Section III, we provide numerical evidence of its reliability.

\section{Numerical results and discussion}

In this section we compare the models previously described to numerical micromagnetic simulations in order to validate them. Radial spin waves with a spiral profile from the core region can also propagate, when the excitation signal is applied in-plane of the patch \cite{article}. This effect can be added to the model in terms of a normalised azimuthal factor that creates a counterclockwise spiralling effect as observed in the simulated wavefronts ($\lambda(\theta) \rightarrow \frac{\theta}{2\pi}\lambda(\theta)$). In micromagnetic simulations, a continuous out-of-plane wave excitation of frequency $\omega_{0}/2\pi= 8.8$ GHz is applied to the core region. The excitation frequency is chosen so the radial spin wave shows a clear wavefront for the given dimensions and material of the magnetic patch \cite{PhysRevB.100.214437}). From Fig. 1 (and numerical results from \cite{PhysRevB.100.214437}), the wavelengths of the spin wave into the domain and the confined mode are chosen as $\lambda_{\text{SSW}} = 135$ nm and $\lambda_{\text{WM}} = 89$ nm. The condition $\lambda_{\text{SSW}}>\lambda_{\text{WM}}$ is satisfied by applying an oscillating magnetic field in the GHz range. We need to address that the main objective is to test the relative change between these two wavelengths, regardless of their absolute values. Fig. 5(a) shows snapshots of the dynamic out-of-plane magnetisation from micromagnetic simulations of two different shapes, a 'double' teardrop shape (two vertices) and a square (four vertices). Their respective k-space maps from each image are shown on the right, where the white arrows indicate the propagation of the main modes, surface spin-waves and Winter's magnons. The images are interpolated for clarity. Before performing a spatial FFT, a Hamming window of 256 points is applied to the data set to avoid image artifacts due to reflections at the edges and spurious high frequency values.

\begin{figure}[ht]
\centering 
\includegraphics[trim=0cm 0cm 0cm 0cm, clip=true, width=9cm]{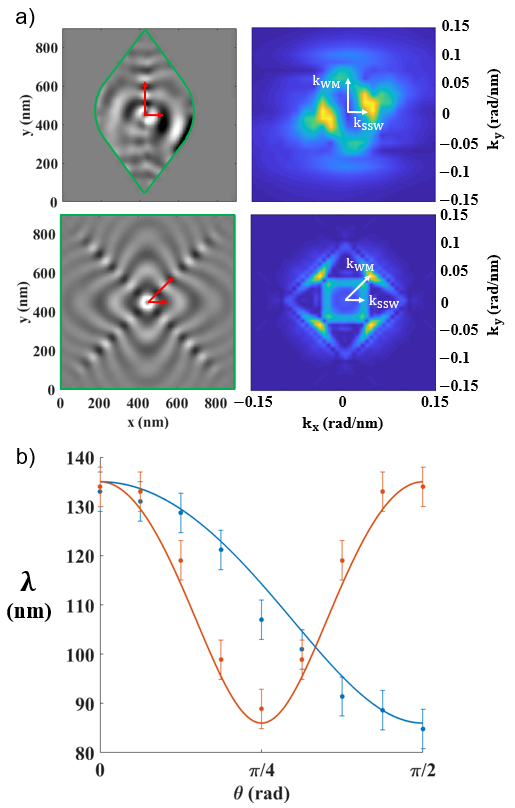}
\caption{(a) (Left) Snapshots of numerical results for a \lq double' teardrop shape ($n=2$) and a square ($n=4$). Red arrows indicate the direction of propagation of the two main modes (surface spin wave and Winter's magnon). (Right) k-space of the snapshots where the wavenumber profile (inverse of the wavelength profile) is shown. Results are interpolated to 5 extra points between data points for clarity. White arrows show the direction of propagation of the main modes. (b) Comparison of the maximum values of $\lambda=2\pi/k$ found in (a) for the lq double' teardrop shape (blue) and the square (orange) with the results from Eq. (6) where $\lambda_{\text{SSW}}=135$ nm and $\lambda_{\text{WM}}=89$ nm found from numerical results at $\omega_{0}/2\pi = 8.8$ GHz and $p \approx 1.75$. Error bars are found after the interpolation process, which introduces a measured error of approximately 8 nm.}  \label{Fields1}
\end{figure}

\par Values of wavelength are extracted from the simulated k-space images at angles from $0$ rad to $\pi/2$ in the teardrop shape and from $0$ rad to $\pi/4$ in the square in steps of $\pi/16$. The error bars are found after interpolation, yielding an error in wavelength of approximately 8 nm. The change in wavelength given by the model shows very good agreement with numerical results and follow the predicted trend. The analytical results can easily be extrapolated to any n-corner shape. Taking this into account, we can confidently say that the proposed model describes the spin wave wavefront of an emitted spin wave from the core in any n-corner shape with accuracy. \par Finally, in any n-corners shape presenting angular periodicity, the azimuthal transition from a surface spin wave of wavelength $\lambda_{1}$ into that of a confined mode along a domain wall $\lambda_{2}<\lambda_{1}$ is of the form of an Hippopede curve or, more generalised,

\begin{gather}
\lambda(\lambda_{1},\lambda_{2},\theta,n)=\lambda_{1}f(\lambda_{1},\lambda_{2},\theta,n)\sqrt{1-g(\lambda_{1},\lambda_{2})\text{cos}^2(\theta\frac{n}{2})}, 
\end{gather}
where functions $f$ and $g$ are generic functions of the indicated magnitudes.
\par As a suggested improvement to the model, a radial dependence could be included into Eq. (7).The radial dependence should consider that, as $n$ increases, the domain walls will merge closer to the core region, which implies no azimuthal gradient and the wavefront profile will be that of a disk (equivalent to $n=0$) under a critical effective radius.
\par The Hippopede curves are obtained from Eq. (7) under the condition: $p>>1$. It is worth noting that the condition $\sqrt{2}\lambda_{\text{WM}}>\lambda_{\text{SSW}}>\lambda_{\text{WM}}$, which implies an \lq hippopede wavefront' with geometrical parameters $b<a$, is not necessary satisfied for all values of $\lambda(\theta)$ at every $\omega_{0}$, as explained before. Although the extended model avoids an upper frequency bound, there is a lower frequency bound from which only spin waves will be radiated into the domains from the core due to the non-zero internal field there. In contrast, a gapless mode can propagate in the domain walls \cite{PhysRevLett.114.247206}. As $\omega_{0}$ increases, at the limit when $\lambda_{\text{SSW}}\approx \lambda_{\text{WM}}$, the wavefront tends to a circular profile, as explained elsewhere \cite{PhysRevB.100.214437} and as Eq. (7) consistently predicts. This frequency dependence is already implicit in the model in parameter $p=\omega_{M}/2\omega_{0}$. At high frequencies where $\omega_{0}\approx\omega_{M}/2$ and therefore $p\approx 1$ (so $p>>1$ is not hold), assuming $\lambda_{\text{SSW}} \approx \lambda_{\text{WM}}$, Eq. (7) effectively yields a wavelength of $\lambda(\theta,n)=\lambda_{\text{SSW}}$, this is, a circular wavefront, which is confirmed by micromagnetic simulations and elsewhere. 
\par At even higher frequencies ($\omega_{0}>>\omega_{M}$), $p<<1$ and therefore $(1-p)/(\text{cos}^2(n\theta/2)-p)\approx 1/\text{cos}^2(n\theta/2)$, Eq. (7) leads again to $\lambda(\theta,n)=\lambda_{\text{SSW}}$, regardless of $\lambda_{\text{SSW}}$ and $\lambda_{\text{WM}}$ values. This implies that Eq. (7) is still a valid model even for $p\approx 1$ or $p << 1$, or in other words, it does not show an upper frequency bound for radial waves. This is consistent with the physical scenario to describe and previous work on radial propagating spin waves.

\par Previous work on exchange-dominated radial spin waves have predominantly dealt with direct observation or experimental detection, as referred to in section I. We believe that our results may help in obtaining further information of these spin waves such as an expression for a spatial-dependent wavelength, potential detection of magnetic inhomogeneities in magnetic films of various geometries or characterization of the material properties, when used, for example, as a fitting tool. We would also like to highlight the applicability of these mathematical curves itself. We believe that, in addition to their known applications in mechanical linkages (see Supplemental Material (1)), the work proposed here is another interesting use of these curves for modeling in physics and in particular, a novelty in Magnetism.

\section{Summary}
\par We have used geometrical expressions to successfully model propagating spin waves from the vortex core region in n-corners elements in a magnetic flux closure configuration where domain walls are present. The proposed models are validated and all show very good agreement with numerical results. The equations can be generalised to any n-corner shape, including non regular shapes such as a teardrop shape (one corner) and a double teardrop shape (two corners). A first model is based on a special case of the \lq hippopede' curves, known as Booth's ovals, since they allow smooth transitions between two known wavelength values. A more exact model is obtained straight from generalising the fundamental equation of surface spin wave dispersion, which can be retrieved by setting $\lambda_{\text{WM}}=0$. This more compact model describes the spin wave wavefront accurately at positions far from the core and specially, close to the inhomogeneous areas (i.e., domain walls). Interestingly, through algebraic transformations, the final equation of the model (where $\lambda_{\text{WM}}\neq0$ and $p>>1$), is identical to the equation of the \lq hippopede' curve. The result connects the mathematical model with the physical parameters of the material and proves the \lq hippopede curves' as the most accurate mathematical description of the transition between one wavelength and the other. Reciprocally, the magnetic properties of the material (through $\lambda_{\text{WM}}$ and $\lambda_{\text{SSW}}$) can be retrieved from the geometrical parameters of the plotted Hippopede curve ($a$ and $b$). Given the frequency of the oscillating field $\omega_{0}$, parameter $p$ and therefore $\omega_{M}$ and $M_{S}$ can also be retrieved.
\par The model from Eq.(7) also takes into account the frequency dependence of the oscillating field. At lower frequencies, the model yields hippopede curves for the spin wave wavefront profile. At higher frequencies, it effectively leads to circular wavefronts, as expected from numerical results.
\par The models can also be applied on spiral wavefronts, although they are originally defined for ‘in-phase’ wavefronts, which makes them suitable for also describing circular/non-spiralling wavefronts. For modelling spiral wavefronts, the equation must be modified accordingly by simply introducing a normalised spiralling effect factor. We hope these results help to better understand the propagating features of spin waves in confined structures, more specially those emitted from quasi-punctual sources and how to control their dynamical properties.

\section{Acknowledgements}
\par This work was supported by EPSRC and the CDT in Metamaterials, University of Exeter. All data created during this research are openly available from the University of Exeter's institutional repository at https://ore.exeter.ac.uk/repository/ 

\bibliography{library}

\begin{thebibliography}{24}%
\makeatletter
\providecommand \@ifxundefined [1]{%
 \@ifx{#1\undefined}
}%
\providecommand \@ifnum [1]{%
 \ifnum #1\expandafter \@firstoftwo
 \else \expandafter \@secondoftwo
 \fi
}%
\providecommand \@ifx [1]{%
 \ifx #1\expandafter \@firstoftwo
 \else \expandafter \@secondoftwo
 \fi
}%
\providecommand \natexlab [1]{#1}%
\providecommand \enquote  [1]{``#1''}%
\providecommand \bibnamefont  [1]{#1}%
\providecommand \bibfnamefont [1]{#1}%
\providecommand \citenamefont [1]{#1}%
\providecommand \href@noop [0]{\@secondoftwo}%
\providecommand \href [0]{\begingroup \@sanitize@url \@href}%
\providecommand \@href[1]{\@@startlink{#1}\@@href}%
\providecommand \@@href[1]{\endgroup#1\@@endlink}%
\providecommand \@sanitize@url [0]{\catcode `\\12\catcode `\$12\catcode
  `\&12\catcode `\#12\catcode `\^12\catcode `\_12\catcode `\%12\relax}%
\providecommand \@@startlink[1]{}%
\providecommand \@@endlink[0]{}%
\providecommand \url  [0]{\begingroup\@sanitize@url \@url }%
\providecommand \@url [1]{\endgroup\@href {#1}{\urlprefix }}%
\providecommand \urlprefix  [0]{URL }%
\providecommand \Eprint [0]{\href }%
\providecommand \doibase [0]{http://dx.doi.org/}%
\providecommand \selectlanguage [0]{\@gobble}%
\providecommand \bibinfo  [0]{\@secondoftwo}%
\providecommand \bibfield  [0]{\@secondoftwo}%
\providecommand \translation [1]{[#1]}%
\providecommand \BibitemOpen [0]{}%
\providecommand \bibitemStop [0]{}%
\providecommand \bibitemNoStop [0]{.\EOS\space}%
\providecommand \EOS [0]{\spacefactor3000\relax}%
\providecommand \BibitemShut  [1]{\csname bibitem#1\endcsname}%
\let\auto@bib@innerbib\@empty
%</preamble>
\bibitem [{\citenamefont {Kruglyak}\ \emph {et~al.}(2010)\citenamefont
  {Kruglyak}, \citenamefont {Demokritov},\ and\ \citenamefont
  {Grundler}}]{0022-3727-43-26-264001}%
  \BibitemOpen
  \bibfield  {author} {\bibinfo {author} {\bibfnamefont {V.~V.}\ \bibnamefont
  {Kruglyak}}, \bibinfo {author} {\bibfnamefont {S.~O.}\ \bibnamefont
  {Demokritov}}, \ and\ \bibinfo {author} {\bibfnamefont {D.}~\bibnamefont
  {Grundler}},\ }\href {http://stacks.iop.org/0022-3727/43/i=26/a=264001}
  {\bibfield  {journal} {\bibinfo  {journal} {Journal of Physics D: Applied
  Physics}\ }\textbf {\bibinfo {volume} {43}},\ \bibinfo {pages} {264001}
  (\bibinfo {year} {2010})}\BibitemShut {NoStop}%
\bibitem [{\citenamefont {Karenowska}\ \emph {et~al.}(2016)\citenamefont
  {Karenowska}, \citenamefont {Chumak}, \citenamefont {Serga},\ and\
  \citenamefont {Hillebrands}}]{Karenowska2016}%
  \BibitemOpen
  \bibfield  {author} {\bibinfo {author} {\bibfnamefont {A.~D.}\ \bibnamefont
  {Karenowska}}, \bibinfo {author} {\bibfnamefont {A.~V.}\ \bibnamefont
  {Chumak}}, \bibinfo {author} {\bibfnamefont {A.~A.}\ \bibnamefont {Serga}}, \
  and\ \bibinfo {author} {\bibfnamefont {B.}~\bibnamefont {Hillebrands}},\
  }\enquote {\bibinfo {title} {Magnon spintronics},}\ in\ \href {\doibase
  10.1007/978-94-007-6892-5_53} {\emph {\bibinfo {booktitle} {Handbook of
  Spintronics}}},\ \bibinfo {editor} {edited by\ \bibinfo {editor}
  {\bibfnamefont {Y.}~\bibnamefont {Xu}}, \bibinfo {editor} {\bibfnamefont
  {D.~D.}\ \bibnamefont {Awschalom}}, \ and\ \bibinfo {editor} {\bibfnamefont
  {J.}~\bibnamefont {Nitta}}}\ (\bibinfo  {publisher} {Springer Netherlands},\
  \bibinfo {address} {Dordrecht},\ \bibinfo {year} {2016})\ pp.\ \bibinfo
  {pages} {1505--1549}\BibitemShut {NoStop}%
\bibitem [{\citenamefont {Hoffmann}\ and\ \citenamefont
  {Bader}(2015)}]{PhysRevApplied.4.047001}%
  \BibitemOpen
  \bibfield  {author} {\bibinfo {author} {\bibfnamefont {A.}~\bibnamefont
  {Hoffmann}}\ and\ \bibinfo {author} {\bibfnamefont {S.~D.}\ \bibnamefont
  {Bader}},\ }\href {\doibase 10.1103/PhysRevApplied.4.047001} {\bibfield
  {journal} {\bibinfo  {journal} {Phys. Rev. Applied}\ }\textbf {\bibinfo
  {volume} {4}},\ \bibinfo {pages} {047001} (\bibinfo {year}
  {2015})}\BibitemShut {NoStop}%
\bibitem [{\citenamefont {Taurel}\ \emph {et~al.}(2016)\citenamefont {Taurel},
  \citenamefont {Valet}, \citenamefont {Naletov}, \citenamefont {Vukadinovic},
  \citenamefont {de~Loubens},\ and\ \citenamefont
  {Klein}}]{PhysRevB.93.184427}%
  \BibitemOpen
  \bibfield  {author} {\bibinfo {author} {\bibfnamefont {B.}~\bibnamefont
  {Taurel}}, \bibinfo {author} {\bibfnamefont {T.}~\bibnamefont {Valet}},
  \bibinfo {author} {\bibfnamefont {V.~V.}\ \bibnamefont {Naletov}}, \bibinfo
  {author} {\bibfnamefont {N.}~\bibnamefont {Vukadinovic}}, \bibinfo {author}
  {\bibfnamefont {G.}~\bibnamefont {de~Loubens}}, \ and\ \bibinfo {author}
  {\bibfnamefont {O.}~\bibnamefont {Klein}},\ }\href {\doibase
  10.1103/PhysRevB.93.184427} {\bibfield  {journal} {\bibinfo  {journal} {Phys.
  Rev. B}\ }\textbf {\bibinfo {volume} {93}},\ \bibinfo {pages} {184427}
  (\bibinfo {year} {2016})}\BibitemShut {NoStop}%
\bibitem [{\citenamefont {Kakazei}\ \emph {et~al.}(2004)\citenamefont
  {Kakazei}, \citenamefont {Wigen}, \citenamefont {Guslienko}, \citenamefont
  {Novosad}, \citenamefont {Slavin}, \citenamefont {Golub}, \citenamefont
  {Lesnik},\ and\ \citenamefont {Otani}}]{doi:10.1063/1.1772868}%
  \BibitemOpen
  \bibfield  {author} {\bibinfo {author} {\bibfnamefont {G.~N.}\ \bibnamefont
  {Kakazei}}, \bibinfo {author} {\bibfnamefont {P.~E.}\ \bibnamefont {Wigen}},
  \bibinfo {author} {\bibfnamefont {K.~Y.}\ \bibnamefont {Guslienko}}, \bibinfo
  {author} {\bibfnamefont {V.}~\bibnamefont {Novosad}}, \bibinfo {author}
  {\bibfnamefont {A.~N.}\ \bibnamefont {Slavin}}, \bibinfo {author}
  {\bibfnamefont {V.~O.}\ \bibnamefont {Golub}}, \bibinfo {author}
  {\bibfnamefont {N.~A.}\ \bibnamefont {Lesnik}}, \ and\ \bibinfo {author}
  {\bibfnamefont {Y.}~\bibnamefont {Otani}},\ }\href {\doibase
  10.1063/1.1772868} {\bibfield  {journal} {\bibinfo  {journal} {Applied
  Physics Letters}\ }\textbf {\bibinfo {volume} {85}},\ \bibinfo {pages} {443}
  (\bibinfo {year} {2004})},\ \Eprint
  {http://arxiv.org/abs/https://doi.org/10.1063/1.1772868}
  {https://doi.org/10.1063/1.1772868} \BibitemShut {NoStop}%
\bibitem [{\citenamefont {Bailleul}\ \emph {et~al.}(2007)\citenamefont
  {Bailleul}, \citenamefont {H\"ollinger}, \citenamefont {Perzlmaier},\ and\
  \citenamefont {Fermon}}]{PhysRevB.76.224401}%
  \BibitemOpen
  \bibfield  {author} {\bibinfo {author} {\bibfnamefont {M.}~\bibnamefont
  {Bailleul}}, \bibinfo {author} {\bibfnamefont {R.}~\bibnamefont
  {H\"ollinger}}, \bibinfo {author} {\bibfnamefont {K.}~\bibnamefont
  {Perzlmaier}}, \ and\ \bibinfo {author} {\bibfnamefont {C.}~\bibnamefont
  {Fermon}},\ }\href {\doibase 10.1103/PhysRevB.76.224401} {\bibfield
  {journal} {\bibinfo  {journal} {Phys. Rev. B}\ }\textbf {\bibinfo {volume}
  {76}},\ \bibinfo {pages} {224401} (\bibinfo {year} {2007})}\BibitemShut
  {NoStop}%
\bibitem [{\citenamefont {Bailleul}\ \emph {et~al.}(2006)\citenamefont
  {Bailleul}, \citenamefont {H\"ollinger},\ and\ \citenamefont
  {Fermon}}]{PhysRevB.73.104424}%
  \BibitemOpen
  \bibfield  {author} {\bibinfo {author} {\bibfnamefont {M.}~\bibnamefont
  {Bailleul}}, \bibinfo {author} {\bibfnamefont {R.}~\bibnamefont
  {H\"ollinger}}, \ and\ \bibinfo {author} {\bibfnamefont {C.}~\bibnamefont
  {Fermon}},\ }\href {\doibase 10.1103/PhysRevB.73.104424} {\bibfield
  {journal} {\bibinfo  {journal} {Phys. Rev. B}\ }\textbf {\bibinfo {volume}
  {73}},\ \bibinfo {pages} {104424} (\bibinfo {year} {2006})}\BibitemShut
  {NoStop}%
\bibitem [{\citenamefont {Yu}\ \emph {et~al.}(2002)\citenamefont {Yu},
  \citenamefont {Guslienko}, \citenamefont {Novosad}, \citenamefont {Otani},
  \citenamefont {Shima},\ and\ \citenamefont {Fukamichi}}]{YuCore}%
  \BibitemOpen
  \bibfield  {author} {\bibinfo {author} {\bibfnamefont {K.}~\bibnamefont
  {Yu}}, \bibinfo {author} {\bibfnamefont {K.}~\bibnamefont {Guslienko}},
  \bibinfo {author} {\bibfnamefont {V.}~\bibnamefont {Novosad}}, \bibinfo
  {author} {\bibfnamefont {Y.}~\bibnamefont {Otani}}, \bibinfo {author}
  {\bibfnamefont {H.}~\bibnamefont {Shima}}, \ and\ \bibinfo {author}
  {\bibfnamefont {K.}~\bibnamefont {Fukamichi}},\ }\bibfield  {booktitle}
  {\emph {\bibinfo {booktitle} {Physical Review B - PHYS REV B}},\ }\href@noop
  {} {\ \textbf {\bibinfo {volume} {65}} (\bibinfo {year} {2002})}\BibitemShut
  {NoStop}%
\bibitem [{\citenamefont {Ding}\ \emph {et~al.}(2014)\citenamefont {Ding},
  \citenamefont {N~Kakazei}, \citenamefont {Liu}, \citenamefont {Guslienko},\
  and\ \citenamefont {Adeyeye}}]{higher}%
  \BibitemOpen
  \bibfield  {author} {\bibinfo {author} {\bibfnamefont {J.}~\bibnamefont
  {Ding}}, \bibinfo {author} {\bibfnamefont {G.}~\bibnamefont {N~Kakazei}},
  \bibinfo {author} {\bibfnamefont {X.}~\bibnamefont {Liu}}, \bibinfo {author}
  {\bibfnamefont {K.}~\bibnamefont {Guslienko}}, \ and\ \bibinfo {author}
  {\bibfnamefont {A.}~\bibnamefont {Adeyeye}},\ }\bibfield  {booktitle} {\emph
  {\bibinfo {booktitle} {Scientific reports}},\ }\href@noop {} {\ \textbf
  {\bibinfo {volume} {4}},\ \bibinfo {pages} {4796} (\bibinfo {year}
  {2014})}\BibitemShut {NoStop}%
\bibitem [{\citenamefont {Guslienko}\ \emph {et~al.}(2015)\citenamefont
  {Guslienko}, \citenamefont {Kakazei}, \citenamefont {Ding}, \citenamefont
  {Liu},\ and\ \citenamefont {Adeyeye}}]{Guslienko2015GiantMV}%
  \BibitemOpen
  \bibfield  {author} {\bibinfo {author} {\bibfnamefont {K.~Y.}\ \bibnamefont
  {Guslienko}}, \bibinfo {author} {\bibfnamefont {G.~N.}\ \bibnamefont
  {Kakazei}}, \bibinfo {author} {\bibfnamefont {J.}~\bibnamefont {Ding}},
  \bibinfo {author} {\bibfnamefont {X.~M.}\ \bibnamefont {Liu}}, \ and\
  \bibinfo {author} {\bibfnamefont {A.~O.}\ \bibnamefont {Adeyeye}},\ }in\
  \href@noop {} {\emph {\bibinfo {booktitle} {Scientific reports}}}\ (\bibinfo
  {year} {2015})\BibitemShut {NoStop}%
\bibitem [{\citenamefont {Eshbach}\ and\ \citenamefont
  {Damon}(1960)}]{PhysRev.118.1208}%
  \BibitemOpen
  \bibfield  {author} {\bibinfo {author} {\bibfnamefont {J.~R.}\ \bibnamefont
  {Eshbach}}\ and\ \bibinfo {author} {\bibfnamefont {R.~W.}\ \bibnamefont
  {Damon}},\ }\href {\doibase 10.1103/PhysRev.118.1208} {\bibfield  {journal}
  {\bibinfo  {journal} {Phys. Rev.}\ }\textbf {\bibinfo {volume} {118}},\
  \bibinfo {pages} {1208} (\bibinfo {year} {1960})}\BibitemShut {NoStop}%
\bibitem [{\citenamefont {Dieterle}\ \emph {et~al.}(2019)\citenamefont
  {Dieterle}, \citenamefont {F\"orster}, \citenamefont {Stoll}, \citenamefont
  {Semisalova}, \citenamefont {Finizio}, \citenamefont {Gangwar}, \citenamefont
  {Weigand}, \citenamefont {Noske}, \citenamefont {F\"ahnle}, \citenamefont
  {Bykova}, \citenamefont {Gr\"afe}, \citenamefont {Bozhko}, \citenamefont
  {Musiienko-Shmarova}, \citenamefont {Tiberkevich}, \citenamefont {Slavin},
  \citenamefont {Back}, \citenamefont {Raabe}, \citenamefont {Sch\"utz},\ and\
  \citenamefont {Wintz}}]{PhysRevLett.122.117202}%
  \BibitemOpen
  \bibfield  {author} {\bibinfo {author} {\bibfnamefont {G.}~\bibnamefont
  {Dieterle}}, \bibinfo {author} {\bibfnamefont {J.}~\bibnamefont {F\"orster}},
  \bibinfo {author} {\bibfnamefont {H.}~\bibnamefont {Stoll}}, \bibinfo
  {author} {\bibfnamefont {A.~S.}\ \bibnamefont {Semisalova}}, \bibinfo
  {author} {\bibfnamefont {S.}~\bibnamefont {Finizio}}, \bibinfo {author}
  {\bibfnamefont {A.}~\bibnamefont {Gangwar}}, \bibinfo {author} {\bibfnamefont
  {M.}~\bibnamefont {Weigand}}, \bibinfo {author} {\bibfnamefont
  {M.}~\bibnamefont {Noske}}, \bibinfo {author} {\bibfnamefont
  {M.}~\bibnamefont {F\"ahnle}}, \bibinfo {author} {\bibfnamefont
  {I.}~\bibnamefont {Bykova}}, \bibinfo {author} {\bibfnamefont
  {J.}~\bibnamefont {Gr\"afe}}, \bibinfo {author} {\bibfnamefont {D.~A.}\
  \bibnamefont {Bozhko}}, \bibinfo {author} {\bibfnamefont {H.~Y.}\
  \bibnamefont {Musiienko-Shmarova}}, \bibinfo {author} {\bibfnamefont
  {V.}~\bibnamefont {Tiberkevich}}, \bibinfo {author} {\bibfnamefont {A.~N.}\
  \bibnamefont {Slavin}}, \bibinfo {author} {\bibfnamefont {C.~H.}\
  \bibnamefont {Back}}, \bibinfo {author} {\bibfnamefont {J.}~\bibnamefont
  {Raabe}}, \bibinfo {author} {\bibfnamefont {G.}~\bibnamefont {Sch\"utz}}, \
  and\ \bibinfo {author} {\bibfnamefont {S.}~\bibnamefont {Wintz}},\ }\href
  {\doibase 10.1103/PhysRevLett.122.117202} {\bibfield  {journal} {\bibinfo
  {journal} {Phys. Rev. Lett.}\ }\textbf {\bibinfo {volume} {122}},\ \bibinfo
  {pages} {117202} (\bibinfo {year} {2019})}\BibitemShut {NoStop}%
\bibitem [{\citenamefont {Davies}\ \emph {et~al.}(2015)\citenamefont {Davies},
  \citenamefont {Francis}, \citenamefont {Sadovnikov}, \citenamefont
  {Chertopalov}, \citenamefont {Bryan}, \citenamefont {Grishin}, \citenamefont
  {Allwood}, \citenamefont {Sharaevskii}, \citenamefont {Nikitov},\ and\
  \citenamefont {Kruglyak}}]{article}%
  \BibitemOpen
  \bibfield  {author} {\bibinfo {author} {\bibfnamefont {C.}~\bibnamefont
  {Davies}}, \bibinfo {author} {\bibfnamefont {A.}~\bibnamefont {Francis}},
  \bibinfo {author} {\bibfnamefont {A.}~\bibnamefont {Sadovnikov}}, \bibinfo
  {author} {\bibfnamefont {S.}~\bibnamefont {Chertopalov}}, \bibinfo {author}
  {\bibfnamefont {M.}~\bibnamefont {Bryan}}, \bibinfo {author} {\bibfnamefont
  {S.}~\bibnamefont {Grishin}}, \bibinfo {author} {\bibfnamefont
  {D.}~\bibnamefont {Allwood}}, \bibinfo {author} {\bibfnamefont
  {Y.}~\bibnamefont {Sharaevskii}}, \bibinfo {author} {\bibfnamefont
  {S.}~\bibnamefont {Nikitov}}, \ and\ \bibinfo {author} {\bibfnamefont
  {V.}~\bibnamefont {Kruglyak}},\ }\bibfield  {booktitle} {\emph {\bibinfo
  {booktitle} {Physical Review B - Condensed Matter and Materials Physics}},\
  }\href@noop {} {\ \textbf {\bibinfo {volume} {92}} (\bibinfo {year}
  {2015})}\BibitemShut {NoStop}%
\bibitem [{\citenamefont {Whitehead}\ \emph {et~al.}(2017)\citenamefont
  {Whitehead}, \citenamefont {Horsley}, \citenamefont {Philbin}, \citenamefont
  {Kuchko},\ and\ \citenamefont {Kruglyak}}]{PhysRevB.96.064415}%
  \BibitemOpen
  \bibfield  {author} {\bibinfo {author} {\bibfnamefont {N.~J.}\ \bibnamefont
  {Whitehead}}, \bibinfo {author} {\bibfnamefont {S.~A.~R.}\ \bibnamefont
  {Horsley}}, \bibinfo {author} {\bibfnamefont {T.~G.}\ \bibnamefont
  {Philbin}}, \bibinfo {author} {\bibfnamefont {A.~N.}\ \bibnamefont {Kuchko}},
  \ and\ \bibinfo {author} {\bibfnamefont {V.~V.}\ \bibnamefont {Kruglyak}},\
  }\href {\doibase 10.1103/PhysRevB.96.064415} {\bibfield  {journal} {\bibinfo
  {journal} {Phys. Rev. B}\ }\textbf {\bibinfo {volume} {96}},\ \bibinfo
  {pages} {064415} (\bibinfo {year} {2017})}\BibitemShut {NoStop}%
\bibitem [{\citenamefont {Mushenok}\ \emph {et~al.}(2017)\citenamefont
  {Mushenok}, \citenamefont {Dost}, \citenamefont {Davies}, \citenamefont
  {Allwood}, \citenamefont {Inkson}, \citenamefont {Hrkac},\ and\ \citenamefont
  {Kruglyak}}]{doi:10.1063/1.4995991}%
  \BibitemOpen
  \bibfield  {author} {\bibinfo {author} {\bibfnamefont {F.~B.}\ \bibnamefont
  {Mushenok}}, \bibinfo {author} {\bibfnamefont {R.}~\bibnamefont {Dost}},
  \bibinfo {author} {\bibfnamefont {C.~S.}\ \bibnamefont {Davies}}, \bibinfo
  {author} {\bibfnamefont {D.~A.}\ \bibnamefont {Allwood}}, \bibinfo {author}
  {\bibfnamefont {B.~J.}\ \bibnamefont {Inkson}}, \bibinfo {author}
  {\bibfnamefont {G.}~\bibnamefont {Hrkac}}, \ and\ \bibinfo {author}
  {\bibfnamefont {V.~V.}\ \bibnamefont {Kruglyak}},\ }\href {\doibase
  10.1063/1.4995991} {\bibfield  {journal} {\bibinfo  {journal} {Applied
  Physics Letters}\ }\textbf {\bibinfo {volume} {111}},\ \bibinfo {pages}
  {042404} (\bibinfo {year} {2017})},\ \Eprint
  {http://arxiv.org/abs/https://doi.org/10.1063/1.4995991}
  {https://doi.org/10.1063/1.4995991} \BibitemShut {NoStop}%
\bibitem [{\citenamefont {Stoll}\ \emph {et~al.}(2015)\citenamefont {Stoll},
  \citenamefont {Noske}, \citenamefont {Weigand}, \citenamefont {Richter},
  \citenamefont {Krüger}, \citenamefont {Reeve}, \citenamefont {Hänze},
  \citenamefont {Adolff}, \citenamefont {Stein}, \citenamefont {Meier},
  \citenamefont {Kläui},\ and\ \citenamefont
  {Schütz}}]{10.3389/fphy.2015.00026}%
  \BibitemOpen
  \bibfield  {author} {\bibinfo {author} {\bibfnamefont {H.}~\bibnamefont
  {Stoll}}, \bibinfo {author} {\bibfnamefont {M.}~\bibnamefont {Noske}},
  \bibinfo {author} {\bibfnamefont {M.}~\bibnamefont {Weigand}}, \bibinfo
  {author} {\bibfnamefont {K.}~\bibnamefont {Richter}}, \bibinfo {author}
  {\bibfnamefont {B.}~\bibnamefont {Krüger}}, \bibinfo {author} {\bibfnamefont
  {R.~M.}\ \bibnamefont {Reeve}}, \bibinfo {author} {\bibfnamefont
  {M.}~\bibnamefont {Hänze}}, \bibinfo {author} {\bibfnamefont {C.~F.}\
  \bibnamefont {Adolff}}, \bibinfo {author} {\bibfnamefont {F.-U.}\
  \bibnamefont {Stein}}, \bibinfo {author} {\bibfnamefont {G.}~\bibnamefont
  {Meier}}, \bibinfo {author} {\bibfnamefont {M.}~\bibnamefont {Kläui}}, \
  and\ \bibinfo {author} {\bibfnamefont {G.}~\bibnamefont {Schütz}},\ }\href
  {\doibase 10.3389/fphy.2015.00026} {\bibfield  {journal} {\bibinfo  {journal}
  {Frontiers in Physics}\ }\textbf {\bibinfo {volume} {3}},\ \bibinfo {pages}
  {26} (\bibinfo {year} {2015})}\BibitemShut {NoStop}%
\bibitem [{\citenamefont {Kammerer}\ \emph {et~al.}(2011)\citenamefont
  {Kammerer}, \citenamefont {Weigand}, \citenamefont {Curcic}, \citenamefont
  {Noske}, \citenamefont {Sproll}, \citenamefont {Vansteenkiste}, \citenamefont
  {Waeyenberge}, \citenamefont {Stoll}, \citenamefont {Woltersdorf},
  \citenamefont {Back},\ and\ \citenamefont
  {Schuetz}}]{Kammerer2011MagneticVC}%
  \BibitemOpen
  \bibfield  {author} {\bibinfo {author} {\bibfnamefont {M.}~\bibnamefont
  {Kammerer}}, \bibinfo {author} {\bibfnamefont {M.}~\bibnamefont {Weigand}},
  \bibinfo {author} {\bibfnamefont {M.}~\bibnamefont {Curcic}}, \bibinfo
  {author} {\bibfnamefont {M.}~\bibnamefont {Noske}}, \bibinfo {author}
  {\bibfnamefont {M.}~\bibnamefont {Sproll}}, \bibinfo {author} {\bibfnamefont
  {A.}~\bibnamefont {Vansteenkiste}}, \bibinfo {author} {\bibfnamefont {B.~V.}\
  \bibnamefont {Waeyenberge}}, \bibinfo {author} {\bibfnamefont
  {H.}~\bibnamefont {Stoll}}, \bibinfo {author} {\bibfnamefont
  {G.}~\bibnamefont {Woltersdorf}}, \bibinfo {author} {\bibfnamefont {C.~H.}\
  \bibnamefont {Back}}, \ and\ \bibinfo {author} {\bibfnamefont
  {G.}~\bibnamefont {Schuetz}},\ }in\ \href@noop {} {\emph {\bibinfo
  {booktitle} {Nature communications}}}\ (\bibinfo {year} {2011})\BibitemShut
  {NoStop}%
\bibitem [{\citenamefont {Osuna~Ruiz}\ \emph {et~al.}(2019)\citenamefont
  {Osuna~Ruiz}, \citenamefont {Parra}, \citenamefont {Bukin}, \citenamefont
  {Heath}, \citenamefont {Lara}, \citenamefont {Aliev}, \citenamefont
  {Hibbins},\ and\ \citenamefont {Ogrin}}]{PhysRevB.100.214437}%
  \BibitemOpen
  \bibfield  {author} {\bibinfo {author} {\bibfnamefont {D.}~\bibnamefont
  {Osuna~Ruiz}}, \bibinfo {author} {\bibfnamefont {E.~B.}\ \bibnamefont
  {Parra}}, \bibinfo {author} {\bibfnamefont {N.}~\bibnamefont {Bukin}},
  \bibinfo {author} {\bibfnamefont {M.}~\bibnamefont {Heath}}, \bibinfo
  {author} {\bibfnamefont {A.}~\bibnamefont {Lara}}, \bibinfo {author}
  {\bibfnamefont {F.~G.}\ \bibnamefont {Aliev}}, \bibinfo {author}
  {\bibfnamefont {A.~P.}\ \bibnamefont {Hibbins}}, \ and\ \bibinfo {author}
  {\bibfnamefont {F.~Y.}\ \bibnamefont {Ogrin}},\ }\href {\doibase
  10.1103/PhysRevB.100.214437} {\bibfield  {journal} {\bibinfo  {journal}
  {Phys. Rev. B}\ }\textbf {\bibinfo {volume} {100}},\ \bibinfo {pages}
  {214437} (\bibinfo {year} {2019})}\BibitemShut {NoStop}%
\bibitem [{\citenamefont {Chang}\ \emph {et~al.}(2018)\citenamefont {Chang},
  \citenamefont {Liu}, \citenamefont {Kao}, \citenamefont {Tsai}, \citenamefont
  {Liang},\ and\ \citenamefont {Lee}}]{article_analyticalblochspinwave}%
  \BibitemOpen
  \bibfield  {author} {\bibinfo {author} {\bibfnamefont {L.-J.}\ \bibnamefont
  {Chang}}, \bibinfo {author} {\bibfnamefont {Y.-F.}\ \bibnamefont {Liu}},
  \bibinfo {author} {\bibfnamefont {M.-Y.}\ \bibnamefont {Kao}}, \bibinfo
  {author} {\bibfnamefont {L.-Z.}\ \bibnamefont {Tsai}}, \bibinfo {author}
  {\bibfnamefont {J.-Z.}\ \bibnamefont {Liang}}, \ and\ \bibinfo {author}
  {\bibfnamefont {S.-F.}\ \bibnamefont {Lee}},\ }\href {\doibase
  10.1038/s41598-018-22272-2} {\bibfield  {journal} {\bibinfo  {journal}
  {Scientific Reports}\ }\textbf {\bibinfo {volume} {8}} (\bibinfo {year}
  {2018}),\ 10.1038/s41598-018-22272-2}\BibitemShut {NoStop}%
\bibitem [{\citenamefont {Wang}\ \emph {et~al.}(2013)\citenamefont {Wang},
  \citenamefont {Guo}, \citenamefont {Zhang}, \citenamefont {Nie},\ and\
  \citenamefont {Xia}}]{doi:10.1063/1.4799285}%
  \BibitemOpen
  \bibfield  {author} {\bibinfo {author} {\bibfnamefont {X.-g.}\ \bibnamefont
  {Wang}}, \bibinfo {author} {\bibfnamefont {G.-h.}\ \bibnamefont {Guo}},
  \bibinfo {author} {\bibfnamefont {G.-f.}\ \bibnamefont {Zhang}}, \bibinfo
  {author} {\bibfnamefont {Y.-z.}\ \bibnamefont {Nie}}, \ and\ \bibinfo
  {author} {\bibfnamefont {Q.-l.}\ \bibnamefont {Xia}},\ }\href {\doibase
  10.1063/1.4799285} {\bibfield  {journal} {\bibinfo  {journal} {Applied
  Physics Letters}\ }\textbf {\bibinfo {volume} {102}},\ \bibinfo {pages}
  {132401} (\bibinfo {year} {2013})},\ \Eprint
  {http://arxiv.org/abs/https://doi.org/10.1063/1.4799285}
  {https://doi.org/10.1063/1.4799285} \BibitemShut {NoStop}%
\bibitem [{\citenamefont {DeJong}\ and\ \citenamefont
  {Livesey}(2015)}]{PhysRevB.92.214420}%
  \BibitemOpen
  \bibfield  {author} {\bibinfo {author} {\bibfnamefont {M.~D.}\ \bibnamefont
  {DeJong}}\ and\ \bibinfo {author} {\bibfnamefont {K.~L.}\ \bibnamefont
  {Livesey}},\ }\href {\doibase 10.1103/PhysRevB.92.214420} {\bibfield
  {journal} {\bibinfo  {journal} {Phys. Rev. B}\ }\textbf {\bibinfo {volume}
  {92}},\ \bibinfo {pages} {214420} (\bibinfo {year} {2015})}\BibitemShut
  {NoStop}%
\bibitem [{\citenamefont {Vansteenkiste}\ \emph {et~al.}(2014)\citenamefont
  {Vansteenkiste}, \citenamefont {Leliaert}, \citenamefont {Dvornik},
  \citenamefont {Helsen}, \citenamefont {Garcia-Sanchez},\ and\ \citenamefont
  {Van~Waeyenberge}}]{doi:10.1063/1.4899186}%
  \BibitemOpen
  \bibfield  {author} {\bibinfo {author} {\bibfnamefont {A.}~\bibnamefont
  {Vansteenkiste}}, \bibinfo {author} {\bibfnamefont {J.}~\bibnamefont
  {Leliaert}}, \bibinfo {author} {\bibfnamefont {M.}~\bibnamefont {Dvornik}},
  \bibinfo {author} {\bibfnamefont {M.}~\bibnamefont {Helsen}}, \bibinfo
  {author} {\bibfnamefont {F.}~\bibnamefont {Garcia-Sanchez}}, \ and\ \bibinfo
  {author} {\bibfnamefont {B.}~\bibnamefont {Van~Waeyenberge}},\ }\href
  {\doibase 10.1063/1.4899186} {\bibfield  {journal} {\bibinfo  {journal} {AIP
  Advances}\ }\textbf {\bibinfo {volume} {4}},\ \bibinfo {pages} {107133}
  (\bibinfo {year} {2014})},\ \Eprint
  {http://arxiv.org/abs/https://doi.org/10.1063/1.4899186}
  {https://doi.org/10.1063/1.4899186} \BibitemShut {NoStop}%
\bibitem [{\citenamefont {Winter}(1961)}]{PhysRev.124.452}%
  \BibitemOpen
  \bibfield  {author} {\bibinfo {author} {\bibfnamefont {J.~M.}\ \bibnamefont
  {Winter}},\ }\href {\doibase 10.1103/PhysRev.124.452} {\bibfield  {journal}
  {\bibinfo  {journal} {Phys. Rev.}\ }\textbf {\bibinfo {volume} {124}},\
  \bibinfo {pages} {452} (\bibinfo {year} {1961})}\BibitemShut {NoStop}%
\bibitem [{\citenamefont {Garcia-Sanchez}\ \emph {et~al.}(2015)\citenamefont
  {Garcia-Sanchez}, \citenamefont {Borys}, \citenamefont {Soucaille},
  \citenamefont {Adam}, \citenamefont {Stamps},\ and\ \citenamefont
  {Kim}}]{PhysRevLett.114.247206}%
  \BibitemOpen
  \bibfield  {author} {\bibinfo {author} {\bibfnamefont {F.}~\bibnamefont
  {Garcia-Sanchez}}, \bibinfo {author} {\bibfnamefont {P.}~\bibnamefont
  {Borys}}, \bibinfo {author} {\bibfnamefont {R.}~\bibnamefont {Soucaille}},
  \bibinfo {author} {\bibfnamefont {J.-P.}\ \bibnamefont {Adam}}, \bibinfo
  {author} {\bibfnamefont {R.~L.}\ \bibnamefont {Stamps}}, \ and\ \bibinfo
  {author} {\bibfnamefont {J.-V.}\ \bibnamefont {Kim}},\ }\href {\doibase
  10.1103/PhysRevLett.114.247206} {\bibfield  {journal} {\bibinfo  {journal}
  {Phys. Rev. Lett.}\ }\textbf {\bibinfo {volume} {114}},\ \bibinfo {pages}
  {247206} (\bibinfo {year} {2015})}\BibitemShut {NoStop}%
\end{thebibliography}%

\end{document}